\documentclass{ws-p9-75x6-50}

\def\simlt{\lower.5ex\hbox{$\; \buildrel < \over \sim \;$}}
\def\simgt{\lower.5ex\hbox{$\; \buildrel > \over \sim \;$}} 
\def\gx339{GX339$-$4}
\def\gro1655{GRO~J1655$-$40}

\def\etal{{\it et al}.}

\newcounter{state}
\newcommand{\ion}[2]{\setcounter{state}{#2}#1\,{\footnotesize{\Roman{state}}}}

\begin{document}

\title{The optical emission lines as a probe of 
state transitions in black-hole candidates}

\author{R. Soria}

\address{Mullard Space Science Laboratory, University College London, 
Holmbury St Mary, Dorking, RH5 6NT,
UK\\E-mail: rs1@mssl.ucl.ac.uk}

%%%%%%%%%%%%%%%%%%%%%%%%%%%%%%%%%%%%%%%%%%%%%%%%%%%%%%%%%%%%%%
% You may repeat \author \address as often as necessary      %
%%%%%%%%%%%%%%%%%%%%%%%%%%%%%%%%%%%%%%%%%%%%%%%%%%%%%%%%%%%%%%

\maketitle

\abstracts{Optical spectroscopic studies of emission lines 
in black-hole candidates can help us investigate 
state transitions in those systems. Changes in the 
optical line profiles reflect changes in the geometry 
of the accretion flow, usually associated with X-ray state 
transitions in the inner region. We identify at least four optical 
states in the black-hole candidate GRO J1655$-$40 in outburst, 
and two optical states in GX339$-$4.
}

\section{Optical states in black-hole candidates}
\subsection{Overview}
X-ray state transitions in Galactic black-hole candidates (BHCs) 
are usually explained with changes in the geometry and 
physical conditions of the inflowing matter (which could switch, for example, 
from a keplerian disk to a quasi-spherical inflow). 
See reviews by Tanaka and Lewin~\cite{tl}, Ebisawa, 
Titarchuk and Chakrabarti~\cite{et}, Esin \etal~\cite{en}.
Spectral changes in the X-ray irradiation from the central object 
can, in turn, alter the physical conditions 
in the accreting matter at large distances; 
optical emission lines are a useful probe of 
those regions. X-ray and optical observations 
are therefore complementary for a study of state transitions.

%%%%%%%%%%%%%%%%%%%%%%
\begin{figure}[t]
%\figurebox{22pc}{15pc}{} % to have a box alone
\centerline{\epsfxsize=11.5cm % will enlarge or reduce the postscript figures based on the xsize
\epsfbox{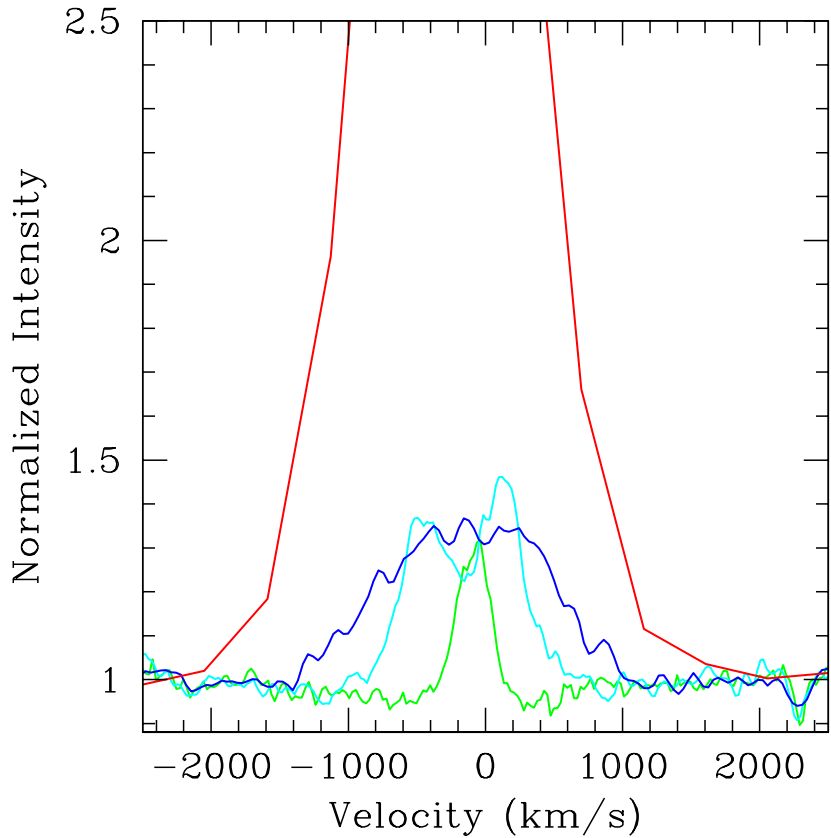}} % postscript image file name
\caption{Different H$\alpha$ emission line profiles corresponding to 
different physical conditions in the accretion flow, in the BHC 
GRO J1655$-$40. 
Dark blue profile: broad, round-topped profile 
(observed on 1996 June 21), interpreted as emission from a disk wind.
Light blue profile: double-peaked profile 
(observed on 1996 June 10), interpreted 
as emission from a geometrically thin, optically thick disk.
Green profile: narrow, single-peaked profile (observed between 1994 August 30 
-- September 4) interpreted as emission from an optically thin, extended 
region above the orbital plane. 
Red profile: the much stronger H$\alpha$ emission line observed 
at low resolution on 1994 
September 6, during a hard X-ray flare. The line is skewed: the extended 
blue wing is evidence of high-velocity outflowing gas.
In all four cases, 
the intensity is normalised to the continuum, and the velocity zeropoint 
is the systemic velocity ($\gamma = -142.4 \pm 1.6$ km s$^{-1}$).
An underlying, broad (FWHM $\simgt 2000$ km s$^{-1}$), shallow  
H$\alpha$ absorption line is also visible in the first 
three cases. \label{fig:fourstates}}
\end{figure}
%%%%%%%%%%%%%%%%%%%%%%%%%%

\subsection{Optical states in outburst}
We carried out high-resolution optical spectroscopic observations 
of the BHC GRO J1655$-$40 in outburst on various occasions 
between 1994 August and 1997 June, using the 3.9m Anglo-Australian Telescope 
and the 2.3m Australian National University telescope 
at Siding Spring Observatory. 
See Soria, Wu \& Hunstead~\cite{sh} and Soria \etal~\cite{ww} 
for a log of our observations and further discussion of our results. 

The optical line profiles
show the existence of a variety of different states in the accretion 
inflow. We classify them into four general states with 
the help of a simple phenomenological model.
The H$\alpha$ line profiles in the four states are shown in 
Figure~1. 
We interpret: 
\begin{itemize}
\item double-peaked emission line profiles 
as emission from a geometrically thin, optically thick accretion disk; 
\item broad, flat-topped 
or round-topped profiles as emission from an optically thick 
($\tau \simgt 1$) wind launched from the disk surface; 
\item strong outburst profiles, observed during hard X-ray flares, 
as emission from an extended optically thick cocoon;
\item  narrow, round-topped profiles, whose width is inconsistent 
with Keplerian rotational velocities in the orbital plane, as emission 
from an unsteady, extended optically thin emitting region 
(cocoon or distribution of clouds) above the disk plane.
\end{itemize}

We have also observed the BHC \gx339 with the 3.9m AAT and 
2.3m ANU telescopes at various times between 1997 May 
and 1999 April (Soria, Wu and Johnston~\cite{sw}; Wu \etal~\cite{ws}). 
Two of the four states found in GRO J1655$-$40 
are similar to the high-soft and low-hard states 
of \gx339. We have not 
found a correspondence for the other two states, which appear to be 
associated with episodes of strong hard X-ray flares and mass ejections.

%%%%%%%%%%%%%%%%
\begin{figure}[h]
%\figurebox{22pc}{15pc}{} % to have a box alone
\centerline{\epsfxsize=10.5cm % will enlarge or reduce the postscript figures based on the xsize
\epsfig{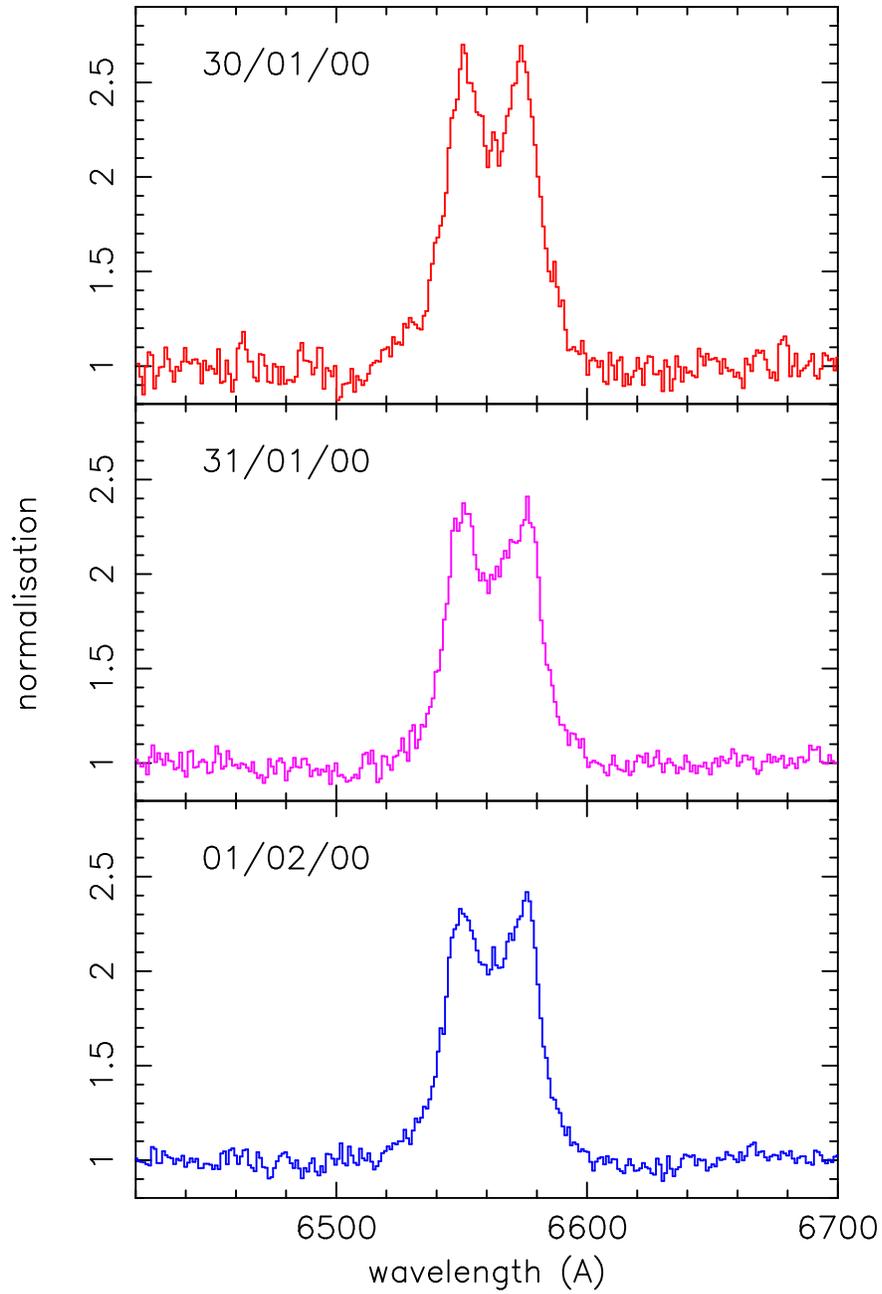}} % postscript image file name
\caption{H$\alpha$ emission line profile observed from the 
BHC A0620$-$00 in quiescence. The data were taken from 
the 2.3m ANU telescope at SSO, on consecutive nights. 
The profile suggests the presence of an optically thin disk 
with a possible contribution from the irradiated face of the secondary.
\label{fig:a0620}}
\end{figure}
%%%%%%%%%%%%%%%%

\subsection{Optical states in quiescence}
The optical spectra of some BHC in quiescence 
show only the signature of the companion star: this is the case of 
\gro1655 and \gx339. 
In other systems, double-peaked optical emission lines 
are always observed even when the X-ray activity 
is very low (corresponding to a low accretion rate). 
The most notable case is A0620$-$00, which we have 
observed with the 2.3m ANU telescope in 2000 January and 2001 January.
The line profile is more symmetric than the disk profiles
observed in outburst. We interpret this ``Smak'' profile 
(Smak \etal~\cite{sk}) as emission from a geometrically 
{\em and} optically thin accretion disk which is always present
around the compact object. A more detailed discussion of this 
result is left to further work.

\section{The four optical states of GRO J1655$-$40}

\subsection{State I: irradiatively-heated thin disk}

\begin{figure}[h]
\vspace{-3.5cm}
\centerline{\epsfig{file=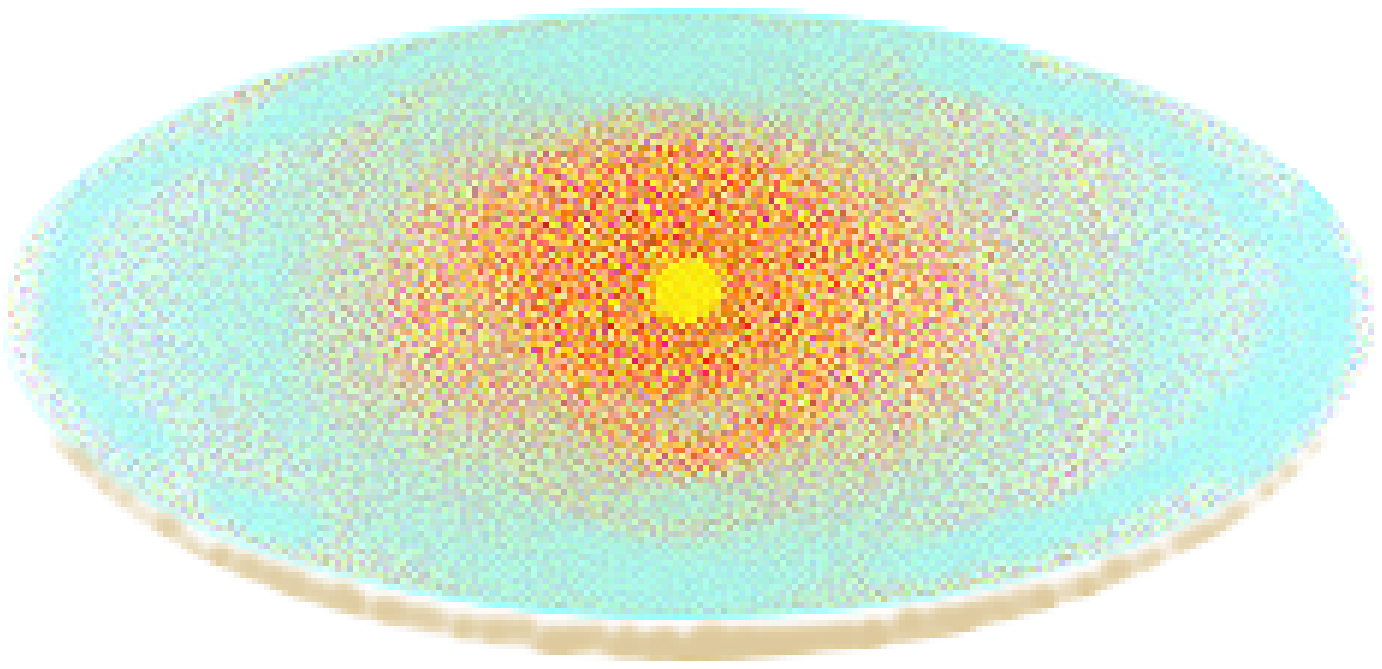,
width=12cm,angle=0}}
\vspace{-2cm}
%\end{figure}
%%%%%%%%%%%%%%%%%%%%%%%

%%%%%%%%%%%%%%%%%%%%%%%
%\begin{figure}
\centerline{\epsfig{file=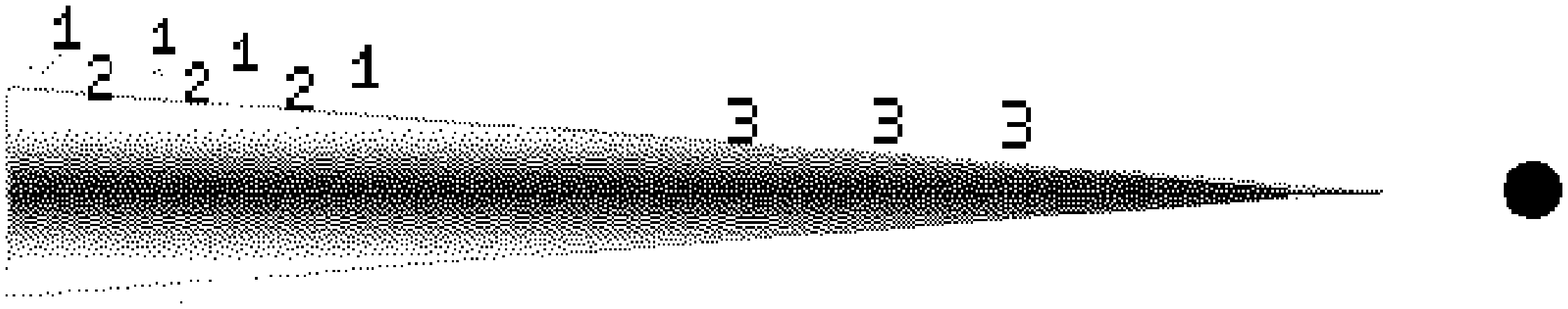,
width=7.5cm,angle=-90}}
\caption{Schematic representation of the accretion disk in the 
high-soft state of GRO J1655$-$40. The disk is geometrically thin, optically 
thick, and irradiated by soft X-rays, which create a temperature-inversion 
layer on its surface, at large radii. \ion{He}{2} (labelled with 1 in the 
cartoon) and Balmer \ion{H}{1} (2)
lines originate in this hot chromosphere. In particular, the Balmer 
emission lines are likely to come from a deeper layer than \ion{He}{2}, 
as their strength is associated with harder irradiation.
At smaller radii, viscous heating is more efficient than X-ray irradiation. 
Broad Balmer absorption lines (3) originate in this region.}

\label{state1}
\end{figure}

This is the ``standard'' state in which an optically thick, 
geometrically thin disk is 
irradiated by the central X-ray source (Figure~\ref{state1}). 
This optical state occurs during 
the high-soft X-ray spectral state. A temperature-inversion layer, 
hotter than the underlying layers, is created at the disk surface owing to 
the irradiation by soft X-rays (e.g. Wu \etal~\cite{ws}). 
The emission lines are double-peaked.  
The velocity separations of the peaks in both the \ion{He}{2} $\lambda 4686$ 
and the H$\alpha$ emission line profiles suggest that the accretion disk 
extends slightly beyond its tidal truncation radius.
Broad absorption lines may come from the inner disk 
where internal viscous heating is more efficient and dominates over
external irradiation. 

\gro1655 was in this state in 1996 May -- early June, 
and throughout most of 1997. When the X-ray irradiation was 
very soft (e.g.\ before 1996 May 27 and in 1997), 
\ion{He}{2} $\lambda 4686$ emission 
was stronger than H$\alpha$ emission. 
The Balmer emission, however, was more prominent when the X-ray 
spectrum was harder (Soria \etal~\cite{sh}). 
The Balmer lines were therefore likely to be emitted 
from deeper layers (at higher densities and lower temperatures) than 
the \ion{He}{2} lines.

This state was also found in \gx339 (Soria \etal~\cite{sw}), and 
in various other BHCs where double-peaked lines were observed.
However, there are differences 
between the high-soft states of \gro1655 and \gx339: (a) 
no broad absorption lines have been observed in \gx339; and (b) \ion{He}{2} 
$\lambda 4686$ was emitted from 
smaller radii than H$\alpha$ in \gx339, while \ion{He}{2} $\lambda 4686$ 
and H$\alpha$ were emitted from similar radii (but at different depths) 
in \gro1655.

%\clearpage
%%%%%%%%%%%%%%%%%%%%%%%
%%%%%%%%%%%%%%%%%%%%%%%
%\clearpage

\subsection{State II: disk outflow}

When a line is emitted in a wind from the disk surface, 
opacity effects tend to reduce the two peaks 
and to increase the intensity of the central trough in the line 
profile. When the line opacity in the wind $\tau_l \simgt 1$, 
the profiles become flat-topped or round-topped (Murray and Chiang~\cite{mc}). 
The blue wing (approaching gas in the outflow) is often more extended 
than the red wing (receding gas, partly shielded from the observer). 
The width of the flat line-top is approximately equal to 
the Keplerian velocity of the disk where the wind is launched.

\begin{figure}[h]
\vspace{-0.5cm}
\centerline{\epsfig{file=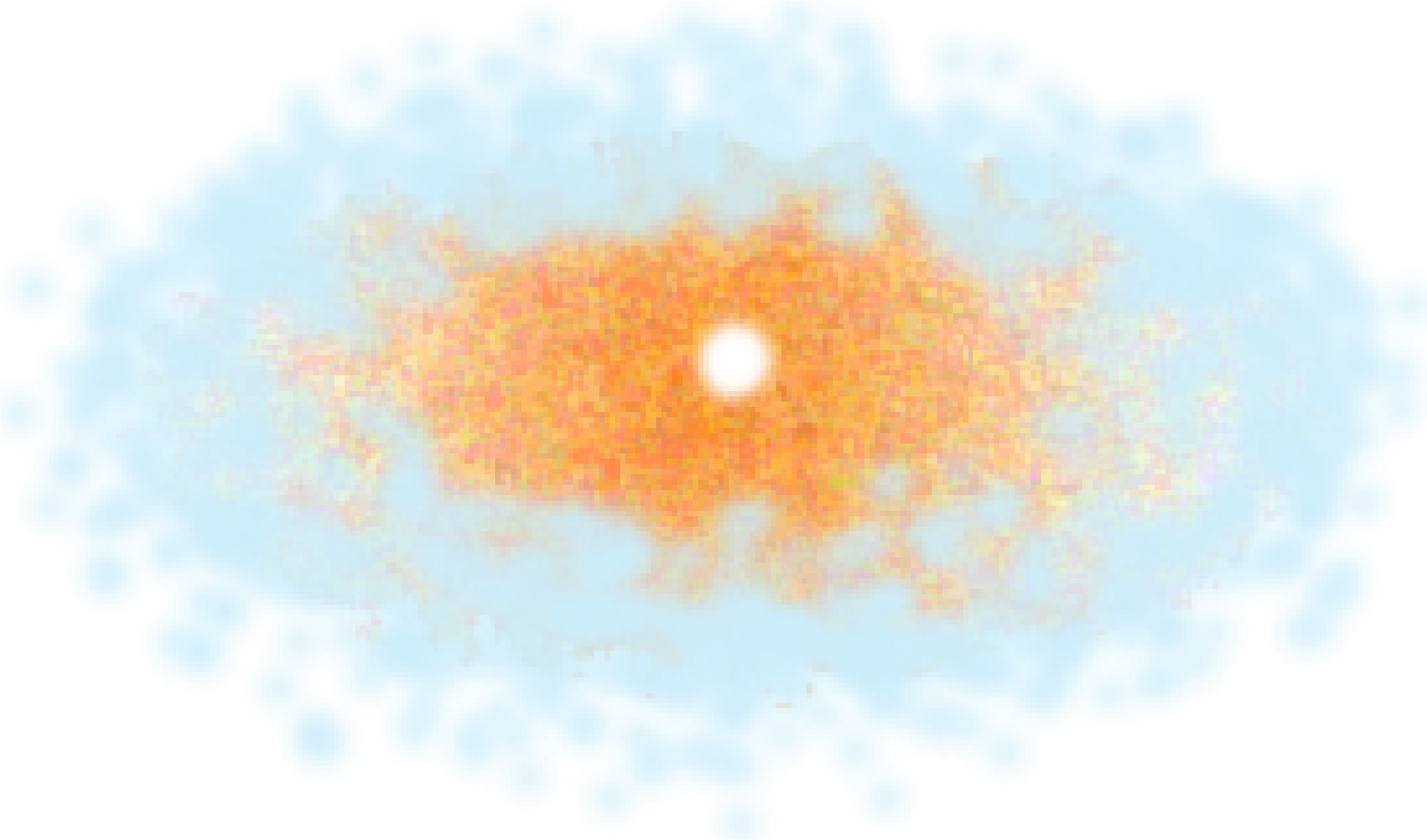,
width=11.8cm,angle=0}}
\vspace{-3cm}
%\end{figure}
%%%%%%%%%%%%%%%%%%%%%%%
%%%%%%%%%%%%%%%%%%%%%%%
%\begin{figure}
\centerline{\epsfig{file=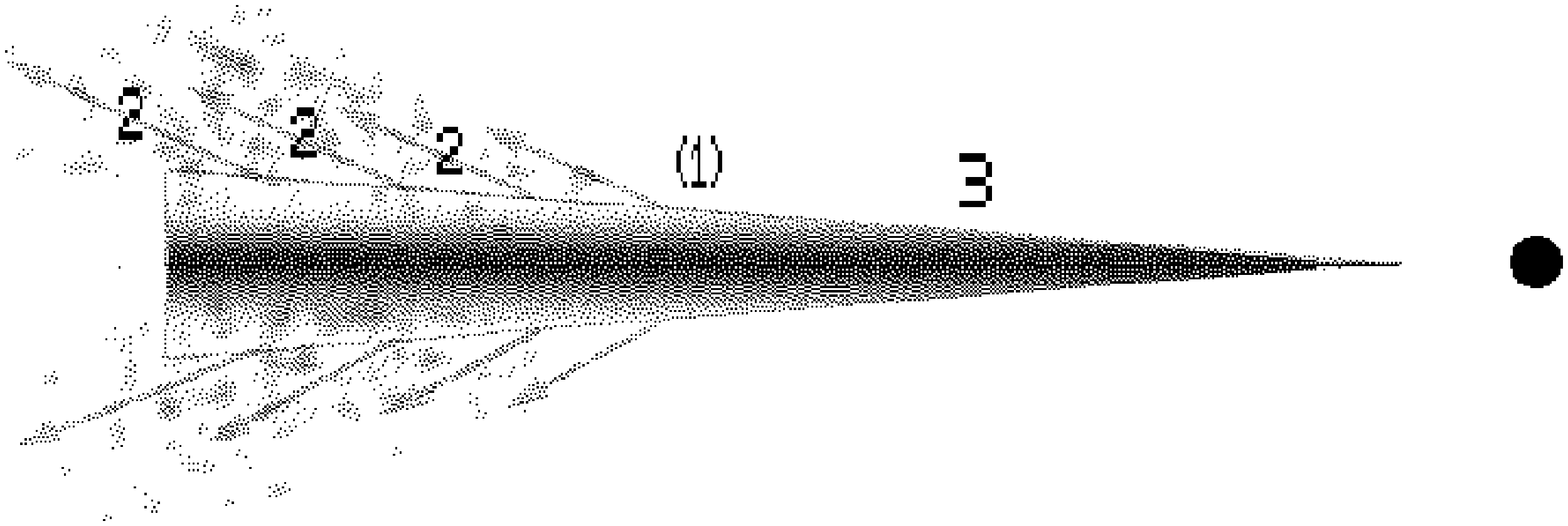,
width=7cm,angle=-90}}
\caption{Schematic representation of the disk-outflow state, 
associated with harder X-ray irradiation. 
As in Figure~\ref{state1}, the label 2 indicates Balmer emission, from an 
outflow region at large radii, and 3 indicates Balmer absorption, from an 
optically-thick disk. In the low-hard state of \gx339, 
\ion{He}{2} $\lambda 4686$ emission (labelled with (1)) was detected 
with a broad, double-peaked profile, 
suggesting that the inner disk was not strongly affected by the wind.}
\label{state2}
\end{figure}

We found evidence of a transition from disk-surface to disk-wind emission 
in \gro1655 on 1996 June 21: the H$\alpha$ profile changed from 
double-peaked to round-topped without substantial changes in its FWHM.
Some evidence of a (weaker) optically-thin wind (e.g.\ 
a more extended blue wing in the emission lines) was also found 
in our June 8 -- 12 spectra.
An increase in the hard X-ray irradiating flux is probably 
responsible for disrupting the geometrically-thin disk and 
driving a wind with substantial Keplerian velocity (Figure~\ref{state2}).

The H$\alpha$ emission line profile in \gx339 also changed from 
double-peaked to 
round-topped with similar FWHM when the system switched from 
a high-soft to a low-hard state (Wu \etal~\cite{ws}). 
In that case, the \ion{He}{2} $\lambda 4686$ 
emission line profile remained double-peaked in the low-hard state, 
suggesting that the inner disk was not strongly affected by the wind.

%%%%%%%%%%%%%%%%%%%%%%%
%%%%%%%%%%%%%%%%%%%%%%%

%\clearpage

\subsection{State III: optically-thick cocoon}

At very high hard X-ray illumination (e.g.\ during the two hard X-ray 
flares observed from \gro1655 in 1994 early-August and early-September) 
there is a further increase in the Balmer line strength. 
The H$\alpha$/H$\beta$ ratio increases dramatically, 
and \ion{H}{1} Paschen lines are seen in emission (Soria \etal~\cite{sh}). 
Higher-ionisation lines such as 
\ion{He}{2} $\lambda 4686$ and the Bowen lines disappear or are greatly 
weakened. The inner-disk broad absorption lines also disappear.

\begin{figure}[h]
\vspace{-1.5cm}
\centerline{\epsfig{file=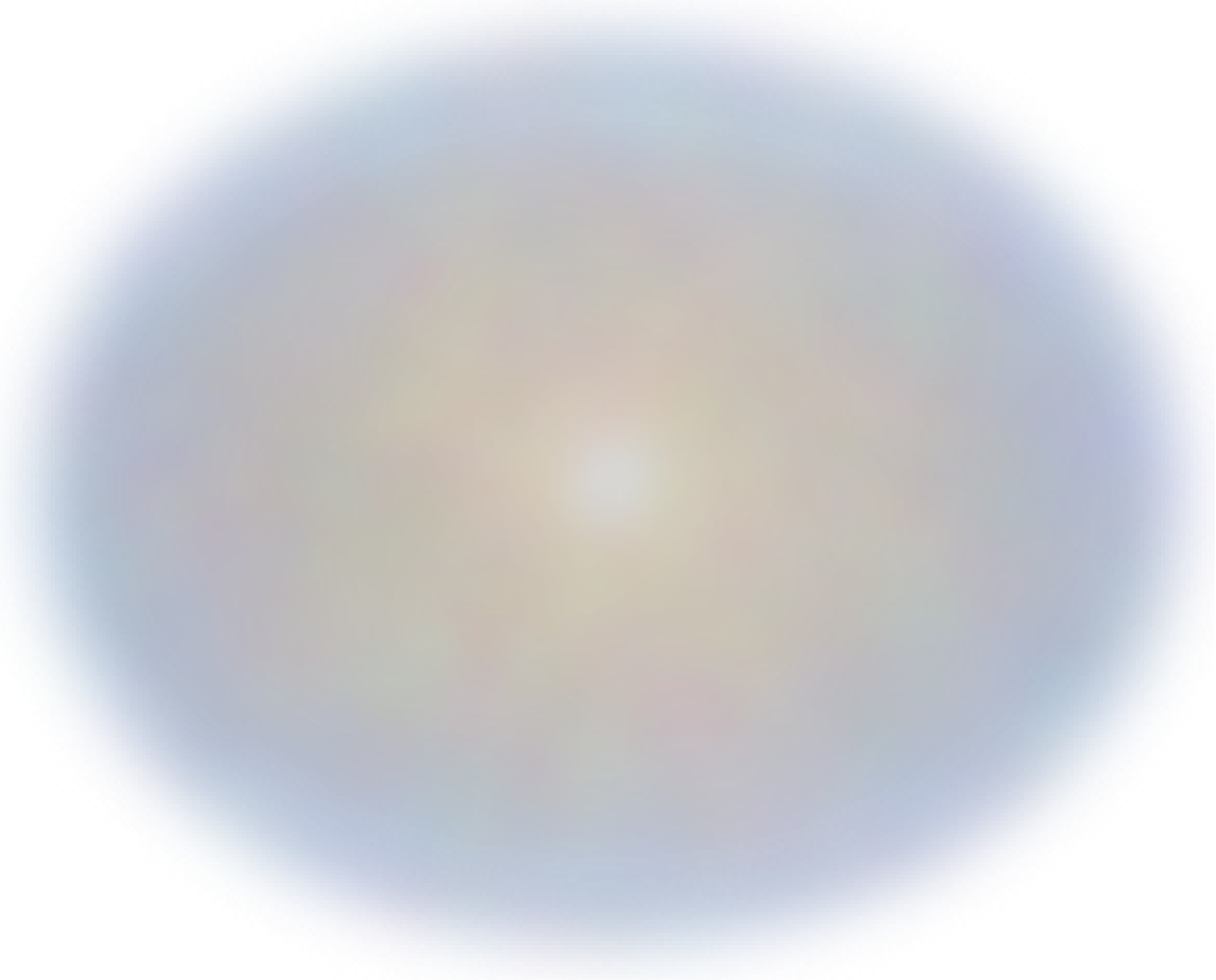,
width=10cm,angle=0}}
\vspace{-1.5cm}
%\end{figure}
%%%%%%%%%%%%%%%%%%%%%%%

%%%%%%%%%%%%%%%%%%%%%%%
%\begin{figure}
\centerline{\epsfig{file=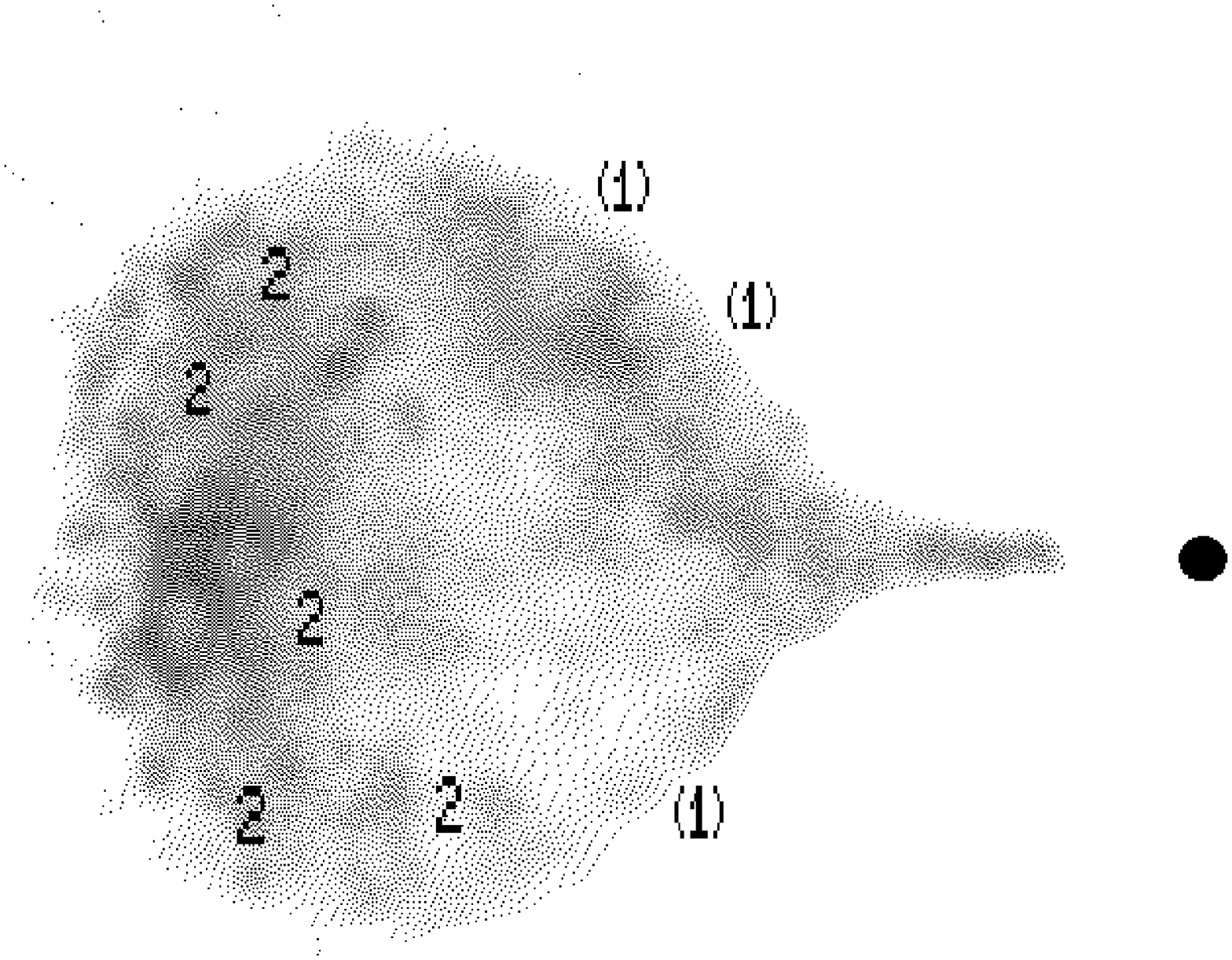,
width=8.5cm,angle=-90}}
\caption{Schematic representation of the accretion flow 
during a hard X-ray flare. No signature of an accretion disk is found. 
The \ion{H}{1} emission (labelled with 2) comes from an extended region 
optically thick to H$\beta$. An extended blue wing in the H$\alpha$ 
emission line is evidence 
of outflowing gas. Higher-ionisation lines (labelled with (1)) 
are weak or absent, probably 
because of geometric occultation of the irradiated face of the gas.}
\label{state3}
\end{figure}

In a hard X-ray outburst, the outer disk can be completely evaporated 
into an optically-thick, extended atmosphere or cocoon because of severe 
X-ray irradiation (Figure~\ref{state3}). 
The disappearance of the higher-ionisation emission 
lines and of the broad absorption at H$\beta$ is probably due to the 
geometric occultation of the inner disk by the outflow (optically thick 
in the H$\beta$/\ion{He}{2} $\lambda 4686$ region).
This is more likely to occur in systems observed at a high orbital 
inclination, such that the inner disk can be occulted by an inflated, 
opaque outer disk region.

\subsection{State IV: optically-thin cocoon}

\begin{figure}[h]
\vspace{-1.5cm}
\centerline{\epsfig{file=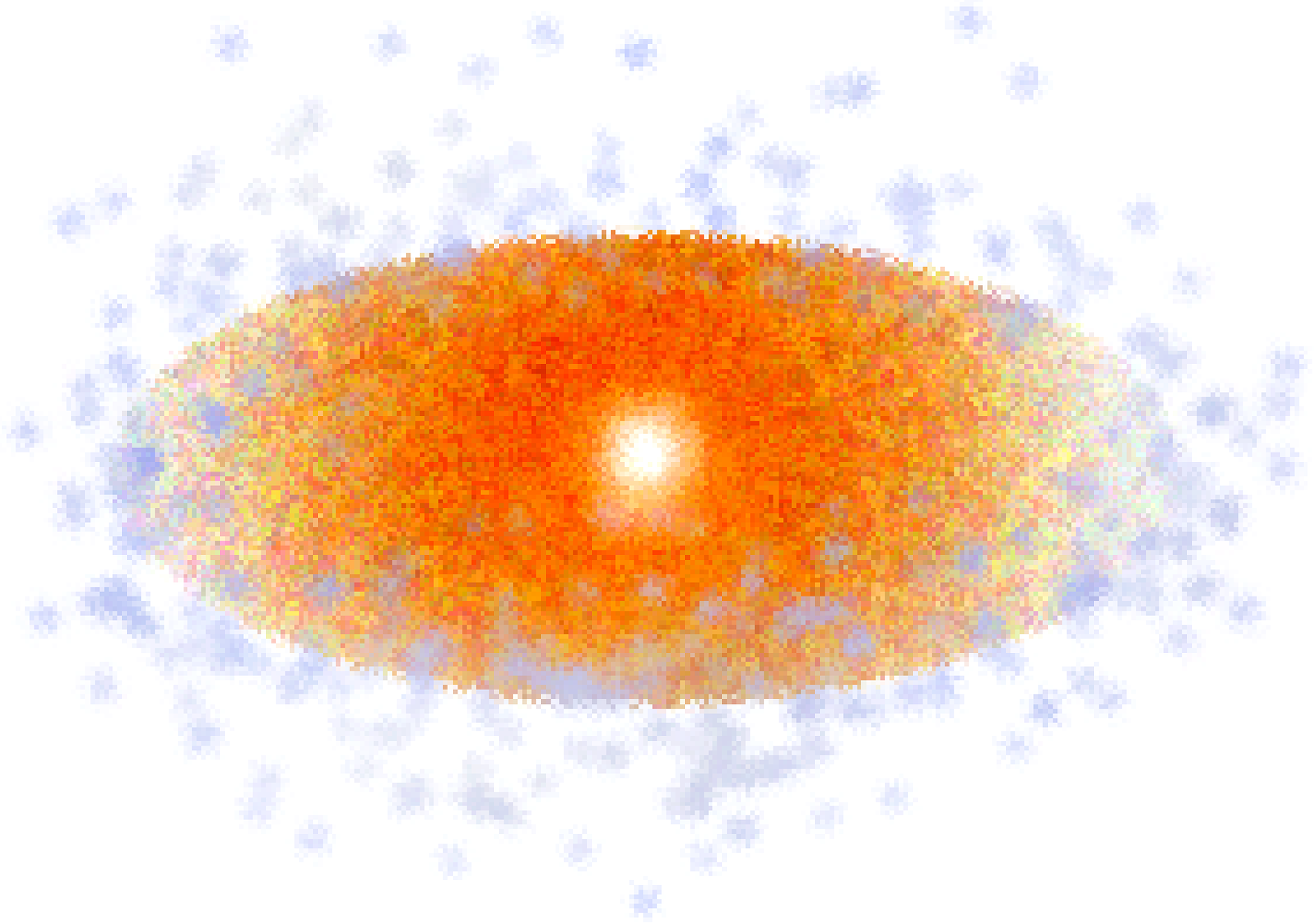,
width=12cm,angle=0}}
\vspace{-2.5cm}
%\end{figure}
%%%%%%%%%%%%%%%%%%%%%%%

%%%%%%%%%%%%%%%%%%%%%%%
%\begin{figure}
\centerline{\epsfig{file=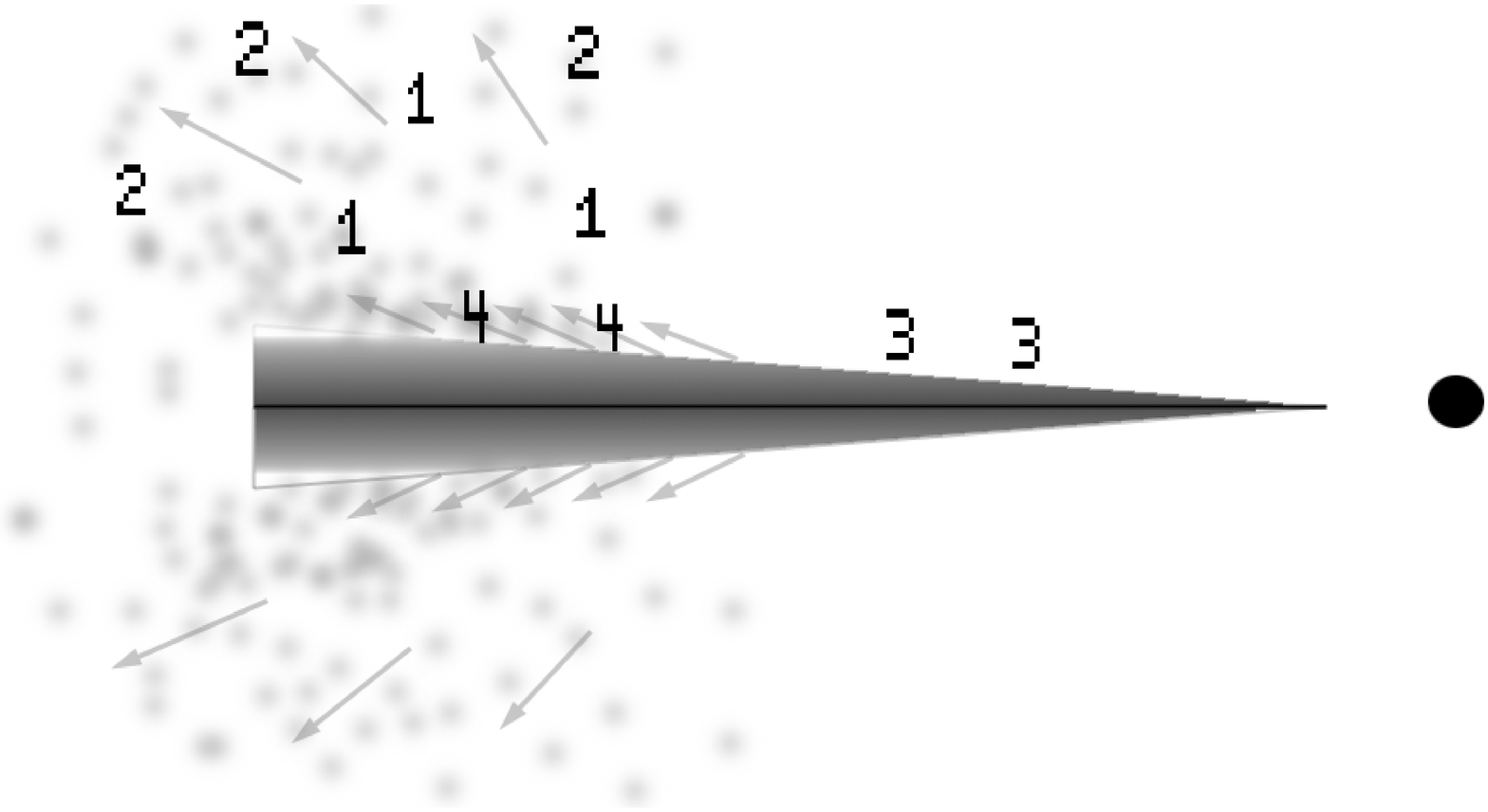,
width=6cm,angle=-90}}
\caption{Schematic representation of the system when an 
extended optically-thin emission region is present. High-ionisation 
He{\footnotesize{II}} and N{\footnotesize{III}} 
Bowen lines (1) and low-ionisation \ion{H}{1} 
Balmer lines (2) are emitted from slow-rotating gas above the disk plane, 
possibly the residue of the earlier mass outflow and disk disruption; 
high-ionisation lines are emitted slightly closer to the disk plane 
and/or at smaller radii. Balmer absorption lines (3) come from the 
optically-thick disk, at small radii. Low-ionisation \ion{N}{2} 
and \ion{O}{2} lines (4) are emitted in a wind from the disk surface 
at large radii.}
\label{state4}
\end{figure}

The optical spectra of \gro1655 obtained two weeks after the major hard 
X-ray flare of 1994 August, and two weeks after the 1994 September flare
look remarkably similar. In both cases, broad absorption lines are 
observed at H$\alpha$ and H$\beta$, suggesting that a disk is present and 
observable (Figure~\ref{state4}).  Broad, flat-topped or 
round-topped emission lines are detected from \ion{N}{2} and \ion{O}{2}, 
suggesting that a wind is launched from the outer disk surface. 
The FWHM of those lines is consistent with the Keplerian velocities 
in the accretion disk.

%%%%%%%%%%%%%%%%%%%%%%%
%%%%%%%%%%%%%%%%%%%%%%%

%%%%%%%%%%%%%%%%%%%%%%%
%%%%%%%%%%%%%%%%%%%%%%%

Strong, narrow emission lines are observed from \ion{H}{1} and \ion{He}{2}. 
Their widths are too narrow to be consistent with Keplerian rotation in the 
disk plane, and their profiles show large, irregular variability between 
a narrow, single-peaked and a slightly broader, double-peaked type. 
The kinematics of the narrow emission lines is consistent with the orbital 
motion of the compact object, but the lines are, on average, blue-shifted 
(signature of expanding gas).

This unsteady state follows episodes of disk 
evaporation and matter ejection. The narrow emission lines are probably due to 
an extended, optically thin, inhomogeneous envelope or distribution 
of gas at large radii, above the disk plane. 
Photoionisation followed by radiative recombination would provide 
a mechanism for the Balmer \ion{H}{1} and \ion{He}{2} emission.
High-inclination systems such as \gro1655 offer the best chance to detect 
line emission from an extended optically-thin region above the disk plane.

\section{Summary}

We have classified the optical spectra of \gro1655, when the system 
is X-ray active, into four general states, 
based on the strength and profile of its emission and absorption lines. 
We showed, with the help of a simple model, the possible 
physical properties and the structure of the accretion inflows and outflows 
in those states.
Two of the four states correspond to the fundamental 
high-soft and low-hard states seen in most transient BHCs 
(for example in \gx339).
The other two states appear to be 
associated with episodes of strong hard X-ray flares 
and mass ejections in the microquasar \gro1655.

\section*{Acknowledgments}
Many thanks to Kinwah Wu, Richard Hunstead and Helen Johnston 
who collaborated in this study.
I would also like to thank Lev Titarchuk for his personal 
support and advice. Most of the work reported here was 
done as a PhD student at the Australian National University. 
I acknowledge financial support from 
University College London and logistical support from 
the Padanian embassy in Rome.

\end{document}